\documentclass[%
preprint,
superscriptaddress,
amsmath,amssymb,
aps, physrev,
]{revtex4-2}

\usepackage{graphicx}
\usepackage{dcolumn}
\usepackage{bm}
\usepackage{hyperref}
\usepackage[dvipsnames]{xcolor}
\usepackage{placeins}

\newcommand{\Bpar}{B_{\!/\!/}}   
\newcommand{\Bperpip}{B_{\perp}^{in}} 
\newcommand{\Bperpout}{B_{\perp}^{out}} 

\begin{document}
	
	
	\title{Gate-tunable anisotropic Josephson diode effect in topological Dirac semimetal Cd$_3$As$_2$ nanowires}
	
	%
	\author{Yan-Liang Hou}
	\affiliation{MESA+ Institute for Nanotechnology, University of Twente, Enschede 7500 AE, The Netherlands}
	
	\author{An-Qi Wang}
	\affiliation{
		State Key Laboratory for Mesoscopic Physics and Frontiers Science Center for Nano-optoelectronics, School of Physics, Peking University, Beijing 100871, China}
        
	\author{Na Li}
	\affiliation{
		State Key Laboratory for Mesoscopic Physics and Frontiers Science Center for Nano-optoelectronics, School of Physics, Peking University, Beijing 100871, China}
        
    \author{Chun-Guang Chu}
	\affiliation{
		State Key Laboratory for Mesoscopic Physics and Frontiers Science Center for Nano-optoelectronics, School of Physics, Peking University, Beijing 100871, China}

    \author{Alexander Brinkman}
	\affiliation{MESA+ Institute for Nanotechnology, University of Twente, Enschede 7500 AE, The Netherlands}
      
	\author{Zhi-Min Liao}
	\affiliation{
		State Key Laboratory for Mesoscopic Physics and Frontiers Science Center for Nano-optoelectronics, School of Physics, Peking University, Beijing 100871, China}
	
	\author{Chuan Li}
	\email{Corresponding author: chuan.li@utwente.nl}
	\affiliation{MESA+ Institute for Nanotechnology, University of Twente, Enschede 7500 AE, The Netherlands}

	%
	%
	
	
	\date{\today}
	
	\begin{abstract}	
	\noindent The intrinsic Josephson diode effect (JDE) has recently attracted considerable attention due to its sensitivity to broken symmetries in Josephson junctions, offering a powerful probe for uncovering hidden symmetry-breaking mechanisms in materials. The presence of higher-harmonic components in the current-phase relation, together with spin-orbital coupling, makes topological materials ideal platforms to explore this effect. In this work, we present a systematic study of the JDE in type-I topological Dirac semimetal Cd$_3$As$_2$ nanowire-based Josephson junctions. We observe a pronounced gate-tunable and highly anisotropic diode response under different magnetic-field orientations. By developing a comprehensive phenomenological model, we capture the angular dependence of the diode effect and, through temperature-dependent measurements, disentangle the respective contributions from bulk and topological surface states. Notably, anomalies in the temperature dependence of the diode efficiency reveal the coexistence of multiple transport channels, highlighting the Josephson diode effect as a sensitive probe of hidden topological superconducting states.
	
\end{abstract}

\maketitle



\section{Introduction}

\noindent The Josephson diode effect (JDE), characterized by a nonreciprocal critical current $|I_c^+| \neq |I_c^-|$, represents a fundamental departure from conventional Josephson transport and enables dissipationless rectification in superconducting circuits. Unlike normal-state diode effects, the JDE is intrinsically phase coherent and therefore highly sensitive to broken symmetries and the microscopic structure of the current--phase relation (CPR). Recent experiments have demonstrated Josephson diode behavior in a broad range of material platforms, including super-lattice\cite{Ando2020}, Rashba semiconductors \cite{Baumgartner2022,Turini2022_InSb}, engineered van der Waals heterostructures \cite{Wu2022_Nb3Br8,Bauriedl2022_NbSe2,deVries2021_Gra,Dez-Mrida2023_Gra,Lin2022_Gra}, and type-II Dirac and Weyl semimetals \cite{Pal2022_NiTe2,Chen2024_MoTe2}, sparking intense interest in both fundamental mechanisms and device-oriented applications (see also the review in Ref.~\cite{Nadeem2023}).

From a symmetry perspective, a dc Josephson diode effect requires the simultaneous breaking of time-reversal symmetry (TRS) and a spatial symmetry that relates forward and backward supercurrents. However, symmetry breaking alone is generally insufficient. As can be understood from the spectrum of Andreev bound states, a non-sinusoidal CPR containing higher harmonics is typically required to lift the equality between positive and negative critical currents \cite{Baumgartner2022,Pal2022_NiTe2}. Microscopically, such CPR asymmetry may arise from finite-momentum Cooper pairing or from the formation of a $\phi_0$-junction, where a phase shift $0 < \phi_0 < \pi$ is induced in the CPR~\cite{Dolcini2015_phi,Yokoyama2014_phi}. When combined with higher-harmonic components, these mechanisms lead to unequal supercurrent amplitudes for the positive and negative branches of the Andreev bound states \cite{Pal2022_NiTe2,Davydova2022_Univ,Lu2023,Zazunov2024,ZhangPRX2022}.

The interplay between superconductivity and topology is particularly appealing in this context. First, topological states with spin--momentum locking can strongly enhance finite-momentum pairing in the presence of TRS breaking \cite{Tanaka2022,Lu2023,Cayao2024}. Second, proximity-induced superconductivity in topological materials can host Majorana zero modes (MZMs), which are key building blocks for topological quantum computation \cite{Fu2008_MZM,Linder2010_MZM}. Recent theoretical works further suggest that parity-protected MZMs can significantly enhance the Josephson diode effect \cite{Legg2023_Parity,Mondal2025,Cayao2024,Wang2024,Picoli2023}. Topological Dirac and Weyl semimetals (DSM and WSM) share many surface-state characteristics with topological insulators while offering a three-dimensional platform to study topological Josephson effects \cite{Crassee2018,Li2021b,Li2020e}.

To date, the Josephson diode effect in topological materials has been predominantly reported in type-II DSM and WSM~\cite{Pal2022_NiTe2,Chen2024_MoTe2}. However, the typical large bulk carrier density in type-II DSM/WSMs limits electrostatic tunability and complicates the disentanglement of surface and bulk contributions. In contrast, Cd$_3$As$_2$ is a type-I Dirac semimetal with ultrahigh mobility and low intrinsic carrier density \cite{Ali2014}, making it an attractive platform for gate-tunable proximity devices and surface-dominated transport \cite{Li2018a,Li2021b,Li2016}. This enables controlled realization of the Josephson diode effect via external fields, spin textures, and electrostatic gating, and provides a unique opportunity to disentangle surface and bulk transport channels \cite{Crassee2018}.

In this work, we report a highly gate-tunable JDE in Cd$_3$As$_2$ nanowire-based Josephson junctions. The diode efficiency reaches a maximum near the Dirac point and exhibits an anomalous enhancement at elevated temperatures around 1.2~K. Notably, the gate and temperature ranges where the JDE is maximized coincide with a transport regime dominated by topological surface states, indicating a strong surface-mediated contribution to the JDE. In addition to large and tunable diode efficiencies, we observe a gate-dependent sign reversal of the diode effect and a pronounced anisotropy with respect to the in-plane magnetic-field orientation. We develop a phenomenological microscopic model incorporating finite-momentum pairing, multi-harmonic CPR components, and interference between surface and bulk channels, which reproduces the main experimental observations and captures the gate and angular dependences of the JDE \cite{ZhangPRX2022,Daido2022,Cayao2024,Mondal2025}.
	
	\section{Results}
	
\begin{figure}[b]
		\includegraphics[width=\columnwidth]{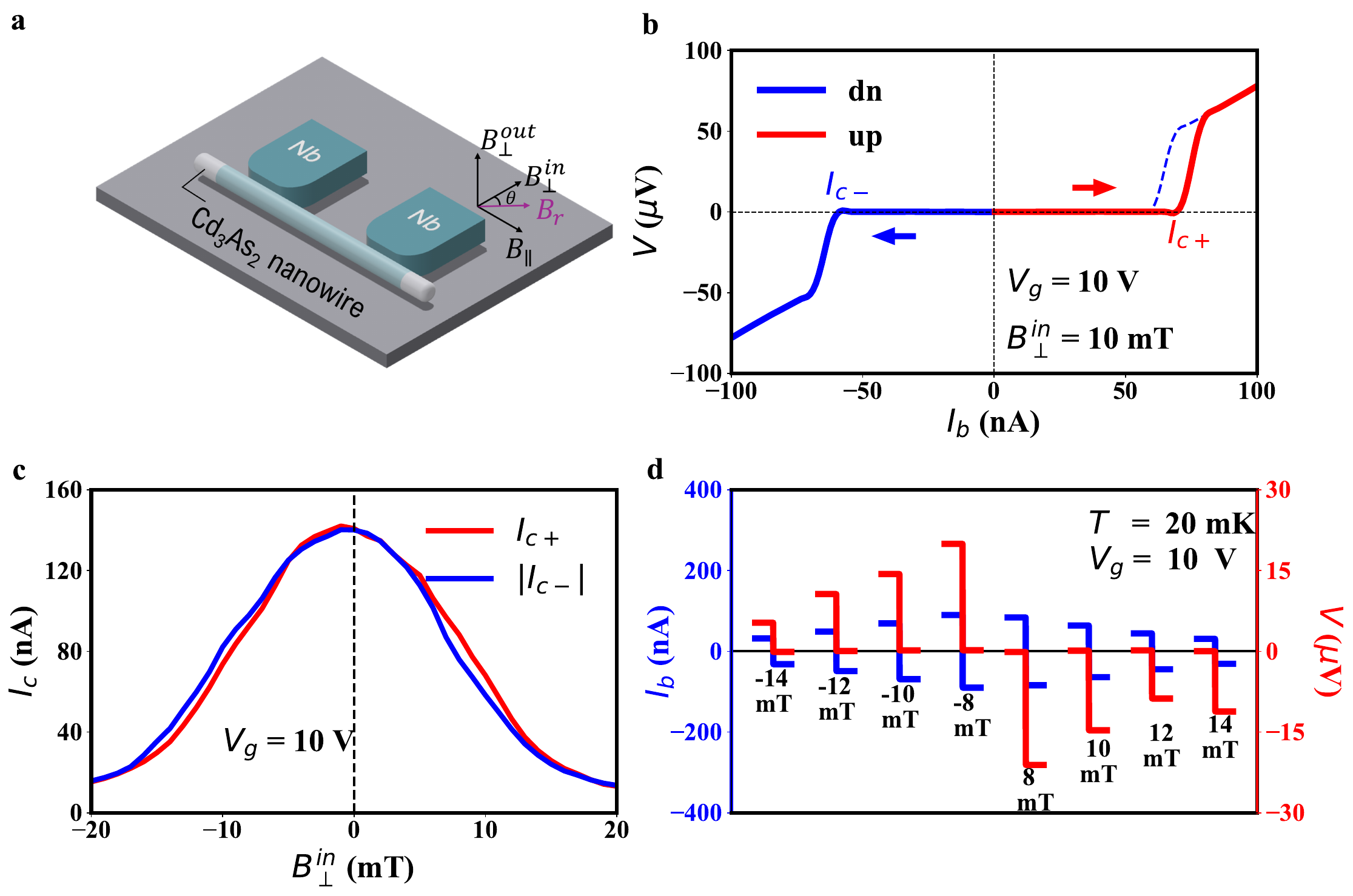}
		\caption{\label{fig:Fig1} 
			\textbf{Josephson diode effect in Cd$_3$As$_2$ (device A: L=800 nm).}  
			\textbf{a}, Schematic of an Nb–Cd$_3$As$_2$ nanowire–Nb Josephson junction. Magnetic field orientations are defined as in-plane perpendicular ($\Bperpip$), in-plane parallel ($\Bpar$), and out-of-plane ($\Bperpout$); the angle $\theta$ is measured relative to $\Bperpip$.  
			\textbf{b}, $I_b$–$V$ characteristics of up- and down-sweeps (starting from 0) at $V_g = 10 \mathrm{V}$ and $\Bperpip = 10~\mathrm{mT}$, showing pronounced nonreciprocal switching currents $I_c^+$ and $I_c^-$. Blue dotted curve: down-sweep data plotted in absolute value.  
			\textbf{c}, Dependence of $I_c^+$ and $\lvert I_c^- \rvert$ on $\Bperpip$ at $V_g = 10~\mathrm{V}$.  
			\textbf{d}, Rectification effect at selected values of $\Bperpip$.
		}
	\end{figure}

	\noindent In this work, we present measurements of three different devices, with nanowire diameters all about 100 nm and junction lengths between 400~nm and 800~nm. Unless otherwise specified, all data were acquired at a base temperature of 20~mK. A schematic of the device and the magnetic field orientations are shown in Fig.~\ref{fig:Fig1}a.

	\renewcommand{\figurename}{Fig.}
    
	\subsection{Josephson diode effect in topological Dirac semimetal}
	
	\noindent Fig.\ref{fig:Fig1}b shows the $I_b$--$V$ characteristics at $V_g=10$~V and $B_y=10$~mT for up-sweeps (red) and down-sweeps (blue, solid line). While taking the absolute value of the negative current (blue dashed line), a clear asymmetry is observed between the critical currents $I_c^+$ and $|I_c^-|$, corresponding to the switching currents in the two sweep directions. If bias current amplitude is set between $I_c^+$ and $|I_c^-|$, the system remains superconducting in one current direction but becomes dissipative in the opposite direction, evidencing a rectification effect, so-called Josephson diode effect.
	
	Fig.~\ref{fig:Fig1}c shows the critical current $I^+_c$ (red) and $|I^-_c|$ (blue) as a function of the in-plane perpendicular magnetic field $\Bperpip$ at $V_g=10$ V. At $\Bperpip=0$, $I^+_c=|I^-_c|$, indicating the preserved TRS. It is clear to see that at the negative field, $\left| I_c^- \right|$ is larger than $ I_c^+ $, and at the positive field, this is reversed, obeying an anti-symmetric dependence of the applied magnetic field. 
	
	This rectification effect is further illustrated in Fig.~\ref{fig:Fig1}d, which shows the voltage response across the junction at $T=20$~mK under selected in-plane magnetic fields $\Bperpip=\pm14$, $\pm12$, $\pm10$, and $\pm8$~mT. The blue curves denote the applied bias currents, while the red curves show the corresponding voltages across the junction. In all cases, the amplitude of the applied bias current lies between $I_c^+$ and $|I_c^-|$, leading to a pronounced rectification response. Moreover, the data reveal a reversal of the rectification polarity depending on the direction of the applied in-plane magnetic field.

	\FloatBarrier
	\subsection{Gate dependence}

	\begin{figure}[hpt]
		\includegraphics[width=1\columnwidth]{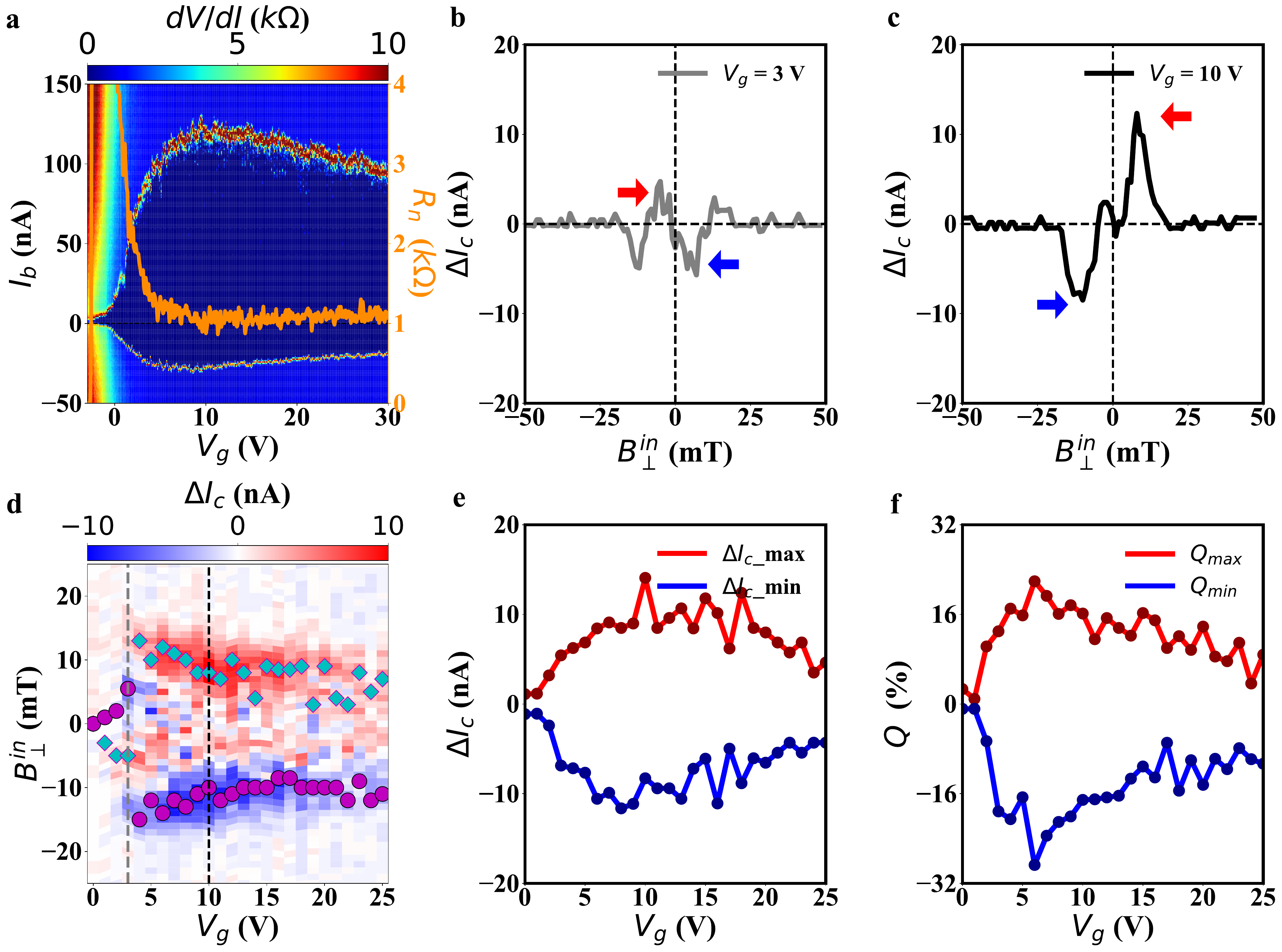}
		\caption{\label{fig:Fig2} 
			\textbf{Gate dependence of the JDE (device A: L=800 nm). }
			\textbf{(a)} Differential resistance $dV/dI$ as a function of gate voltage $V_g$ and bias current $I_b$.  
			The dark blue region corresponds to the superconducting state, which can be continuously tuned and eventually depleted by the gate.  
			The orange curve on the right axis shows the normal-state resistance as a function of $V_g$.  
			\textbf{(b,c)} Dependence of $\Delta I_c$ on the in-plane perpendicular magnetic field ($\Bperpip$) at two representative gate voltages (3 V and 10 V).  
			Blue and red arrows mark the minima and maxima of $\Delta I_c$, respectively.  
			\textbf{(d)} Colormap of $\Delta I_c$ as a function of $V_g$ and $\Bperpip$.  
			Cyan and purple markers indicate the positions of the $\Delta I_c$ maxima and minima in $\Bperpip$ as a function of $V_g$.  
			\textbf{(e,f)} Gate dependence of $\Delta I_c$ and of the Q-factor.
		}
	\end{figure}
	
	\noindent Applying a back-gate voltage strongly tunes the junction properties. In Fig.~\ref{fig:Fig2}a, we show the dV/dI as function of bias current $I_b$ and back-gate voltage $V_g$, where the dark-blue area indicates the superconducting regime. For positive gate voltages, the critical current $I_c$ increases rapidly and peaks near $V_g \approx 11$~V, after which it gradually decreases with further increasing $V_g$. For negative gate voltages, $I_c$ is suppressed to zero within a few volts, indicating depletion of the supercurrent channel. Meanwhile, the normal-state resistance reaches a maximum and remains nearly constant over a wide negative range. A maximum in $I_c$ near the Dirac point has been previously reported in similar systems~\cite{Li2018a,Li2021b}, attributed to an optimized surface-to-bulk contribution: as the Fermi level approaches the Dirac point, bulk carriers diminish more rapidly than topological surface carriers. As a result, the relative weight of surface transport is maximized near the Dirac point, giving rise to the observed maximum in $I_c$.
	
	The JDE is characterized by the critical-current asymmetry $\Delta I_c \equiv I_c^+ - \lvert I_c^- \rvert$ and the efficiency factor $Q \equiv \dfrac{\lvert \Delta I_c \rvert}{(I_c^+ + \lvert I_c^- \rvert)/2}$.	At each gate voltage, we apply an in-plane perpendicular field $\Bperpip$, extract $I_c^+$ and $\lvert I_c^- \rvert$, and compute $\Delta I_c$ and $Q$. Figures~\ref{fig:Fig2}b and \ref{fig:Fig2}c show $\Delta I_c(\Bperpip)$ at $V_g = 3$~V and 10~V. A finite diode signal appears at both polarities of the field. Two key features emerge:
	(1) the sign of $\Delta I_c$ reverses multiple times as the field increase;
	(2) the extrema of $\Delta I_c$, denoted $\Delta I_c^{\mathrm{max}}$ and $\Delta I_c^{\mathrm{min}}$ (blue and red arrows), are tunable by the gate voltage $V_g$.
	
	Previously, multiple sign changes in $\Delta I_c(B)$ were attributed to finite-momentum–induced phase modulation, $\sin(\pi B/B_d)$, where $B_d$ is a characteristic field set by geometry and spin–orbit coupling (SOC) \cite{Pal2022}. Smaller $B_d$ produces more sign changes with field. However, the amplitude of $\Delta I_c$ should be bounded by the superconducting gap $\Delta_s$, which usually follows $\Delta_s \propto \left[1-(|B|/B_c)^2\right]^2$, implying a monotonic decrease in amplitude. This prediction contradicts our data, where the first lobe is smaller than the second. We attribute this unusual behavior to the coexistence of surface and bulk states. The amplitude of $\Delta I_c$ depends on many factors, including junction length, transport regime (ballistic or diffusive), and the resulting higher harmonics in the CPR~\cite{Lu2023}. Topological surface states, with spin–momentum locking, longer coherence lengths, and larger mean free paths~\cite{Li2018a}, naturally yield a stronger JDE contribution. This suggests that the larger lobe in $\Delta I_c$ belongs to the surface states, which corresponds to $B_d^s\approx15$ mT. The inner lobe corresponds to the bulk contribution, corresponding to a smaller $B_d^b\approx5 $mT.
	
	The gate tunability is further illustrated in Fig.~\ref{fig:Fig2}d, where we plot $\Delta I_c$ as a function of both $V_g$ and $\Bperpip$. Figures~\ref{fig:Fig2}b and \ref{fig:Fig2}c correspond to line cuts along the field axis at the gate voltages indicated by dashed lines in Fig.~\ref{fig:Fig2}d. For each gate, we identify $\Delta I_c^{\mathrm{max}}$ and $\Delta I_c^{\mathrm{min}}$ in $\Delta I_c$(B) and trace their evolution in the 2D map using purple and black markers. The field position of the maximum ($B_{\perp,\mathrm{max}}^{\mathrm{in}}$) varies with $V_g$. It remains nearly constant as $V_g$ increases beyond the gate voltage that maximizes $I_c$ ($V_g^{\mathrm{max}} \sim 11$~V). In contrast, when $V_g$ decreases below $V_g^{\mathrm{max}}$, the corresponding $B_{\perp,\mathrm{max}}^{\mathrm{in}}$ first increases, then rapidly drops to zero as the junction gets depleted. This suggests that while tuning the Fermi level, the properties of the states, e.g. effective SOC can be tuned, yielding a larger $B_d$. 
	
	The gate dependence of the $\Delta I_c^{max(min)}$ and the $Q$-factor are shown in Fig.~\ref{fig:Fig2}e and \ref{fig:Fig2}f, respectively. Both quantities exhibit a pronounced maximum near 11~V and fall to zero at lower gate voltages. These distinctive trends are directly connected to the evolution of the topological surface states, further confirming that the observed JDE serves as a sensitive probe of topological transport channels.

	\FloatBarrier
	\subsection{Anisotropic angle dependence of the JDE}
	
	\begin{figure}[hpt]
		\includegraphics[width=1\columnwidth]{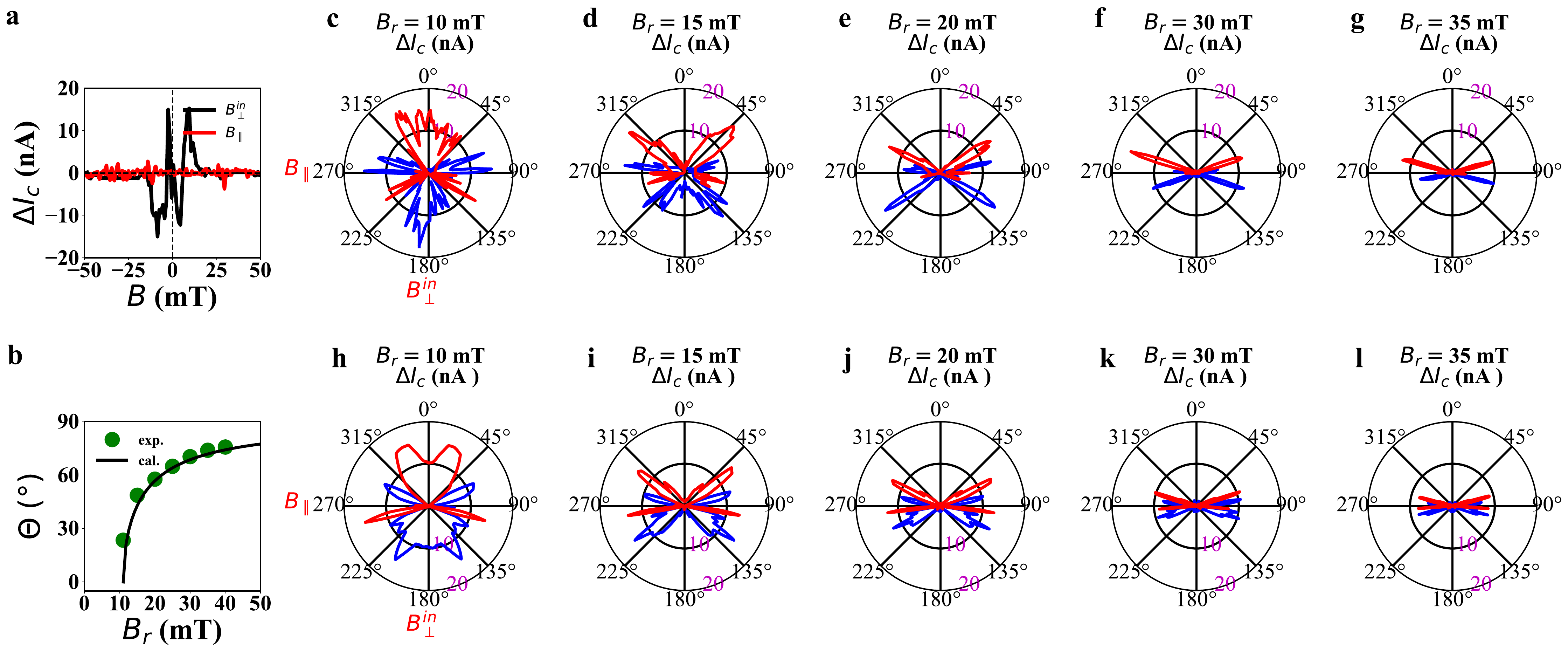}
		\caption{\label{fig:Fig3} 
			In-plane angle dependence of the JDE (device B: L=600 nm).  
			\textbf{(a)} $\Delta I_c$ as a function of $\Bperpip$ (black) and $\Bpar$ (red).  
			\textbf{(b)} Angle corresponding to the maximum $\Delta I_c$ plotted against magnetic field strength $B_r$, comparing experimental data and theoretical calculations.  
			\textbf{(c–g)} Experimental measurements of $\Delta I_c$ versus angle $\theta$ for different magnetic field strengths: $B_r = 10$, 15, 20, 30, and 35 mT. The butterfly-shaped pattern of $\Delta I_c$ in polar coordinates becomes progressively compressed as $B_r$ increases.  
			\textbf{(h–l)} Calculated angular dependence of $\Delta I_c$ for the same set of magnetic field strengths. }
		
	\end{figure}
	
	\noindent To further investigate the observed JDE and the related symmetry-breaking, we rotate the magnetic field in different planes and measure its angle-dependence. 
	First, we investigate the angle dependence in the plane same to the substrate. 
	The experimental results for different field magnitude B$_r$ are shown in Fig.\ref{fig:Fig3}(c-g). In general, the JDE has a mirror symmetry and a two-fold rotation anti-symmetry. At small magnitude (e.g. B$_r$=10mT, Fig.\ref{fig:Fig3}c), the $\Delta I_c$ is maximum along $\Bperpip$ (0\textdegree) direction. When the magnitude $B_r$ is increased, the maximum $\Delta I_c$ is gradually rotated to the $\Bpar$ (90\textdegree) direction, and the lobes in the polar plot become more compressed. Therefore, the JDE becomes highly anisotropic at a higher magnitude of the field. In all cases, the  JDE is anisotropic, and the principle axis of the anisotropy depends on the magnitude of the field.
	
	This behavior mainly arises from the dominant contribution of the in-plane perpendicular field to the JDE. The overall response can be modeled as a linear combination of $\Delta I_c(B_{\perp}^{\mathrm{ip}})$ and $\Delta I_c(B_{\parallel})$. This distinction is clearly demonstrated in Fig.~\ref{fig:Fig3}(a), where the $\Delta I_c$ values for both $B_{\perp}^{\mathrm{ip}}$ and $B_{\parallel}$ are plotted on the same scale. Notably, the $\Delta I_c$ under a parallel field is practically negligible compared to that under the perpendicular field.
	
	To qualitatively understand the angle dependence, we construct a $\Delta I_c$ polar plot based on the results in Fig.~\ref{fig:Fig3}(a), by defining $\Delta I_c(\theta) = \Delta I_c(B_{\perp}^{\mathrm{ip}}(\theta))$, where $B_{\perp}^{\mathrm{ip}}(\theta)$ is obtained directly from the measurement as $B_{\perp}^{\mathrm{ip}}(\theta) = B_r \cos\theta$. In Fig.~\ref{fig:Fig3}(h–l), the calculated angular dependence at different magnitudes of $B_r$ shows good agreement with the experimental results. For direct comparison, the angle corresponding to the maximum $\Delta I_c$ is plotted as a function of $B_r$ in Fig.~\ref{fig:Fig3}(b) for both experiment and calculation. The excellent agreement validates the linear combination model of the two field components in describing the JDE, which can also be extended to other angular planes.

	\begin{figure}[hpt]
		\includegraphics[width=1\columnwidth]{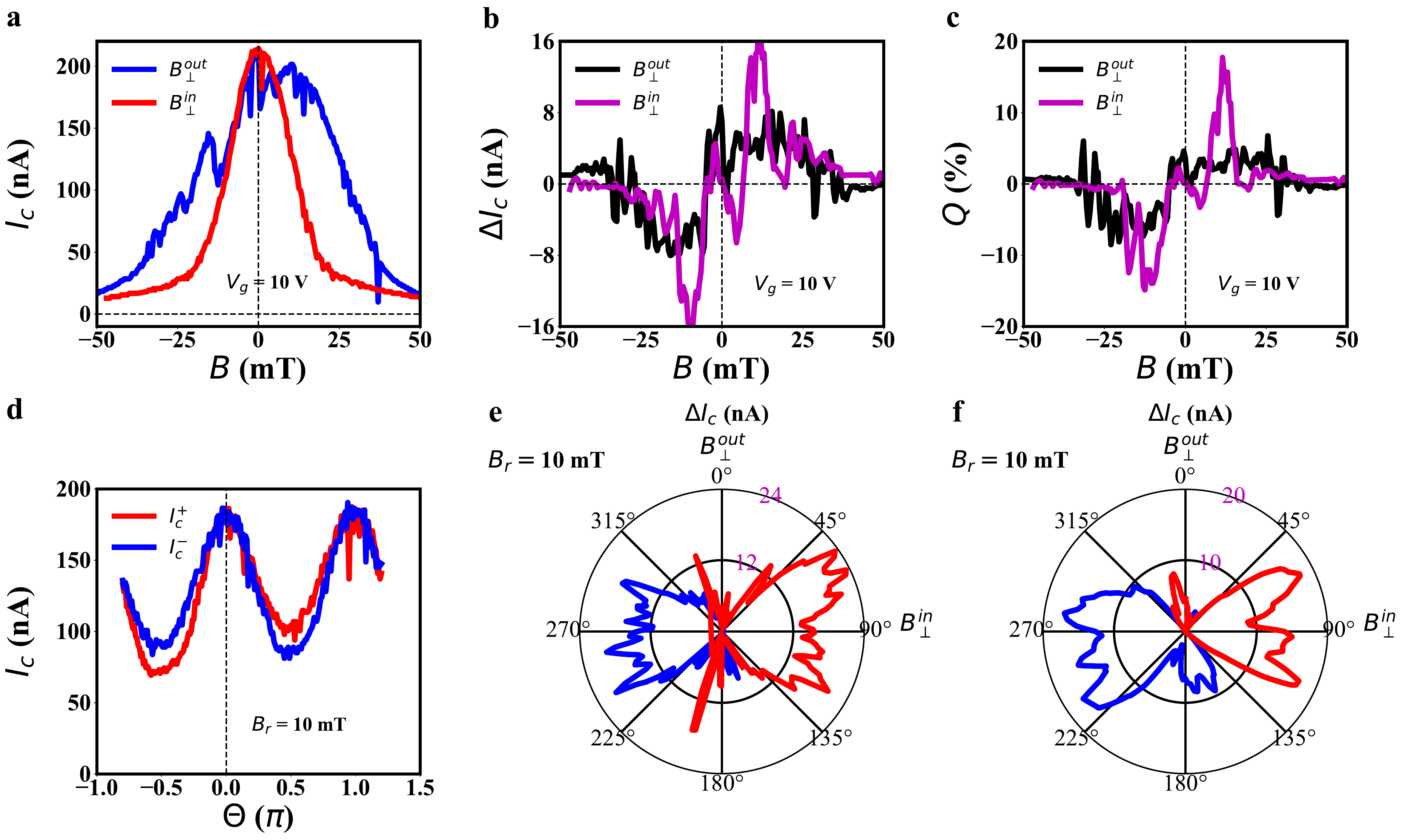}
		\caption{\label{fig:Fig4}
			\textbf{Angle dependence of the JDE in the plane normal to the wire (device C: L=500 nm).}  
			\textbf{(a)} Magnetic-field dependence of the critical supercurrent $I_c$ at $V_g = 10\;\mathrm{V}$. The value of $I_c$ decreases under both in-plane perpendicular field $\Bperpip$ (red) and out-of-plane perpendicular field $\Bperpout$ (blue).  
			\textbf{(b,c)} $\Delta I_c$ and the $Q$-factor as functions of magnetic field strength.  
			\textbf{(d)} Angular dependence of $I_c^{+}$ and $I_c^{-}$ at selected magnetic field strength ($B_r=$ 10 mT). 
            \textbf{(e)}Angular dependence of $\Delta I_c$ at selected magnetic field strength ($B_r=$ 10 mT).
            \textbf{(f)} Calculated angular dependence of $\Delta I_c$ at selected magnetic field strength ($B_r=$ 10 mT), using the results from \textbf{(b)}. 
            }
	\end{figure}
	
	An anisotropic JDE is also observed in the plane perpendicular to the wire. 
	As shown in Fig.~\ref{fig:Fig4}a, the critical supercurrent $I_c$ decreases at different rates under applied magnetic fields in the out-of-plane ($\Bperpout$) and in-plane ($\Bperpip$) directions. Two notable observations can be made:  
	First, contrary to the intuitive expectation that $I_c(\Bperpout)$ would decay faster than $I_c(\Bperpip)$ due to magnetic focusing, $I_c(\Bperpip)$ exhibits a more rapid decrease up to approximately 15~mT, followed by a slower decay toward the critical field. 
	Second, both $\Delta I_c$ and the $Q$-factor oscillate around zero in both directions (~Fig.\ref{fig:Fig4}b and c). Such oscillating behavior can also be seen in the angle dependence at different field strengths (Fig.\ref{fig:Fig4} d).
	Given the wire’s nearly square cross-section, we attribute these features to anisotropic spin–orbit coupling or an anisotropic $g$-factor in the material~\cite{Cao2015, Liu2014, Jeon2014}. A quantitative determination of the SOC strength or the $g$-factor would require a dedicated experiment, which we leave for future work.

	\FloatBarrier
	\subsection{Temperature and length dependence}
	\begin{figure}[hpt]
		\includegraphics[width=1\columnwidth]{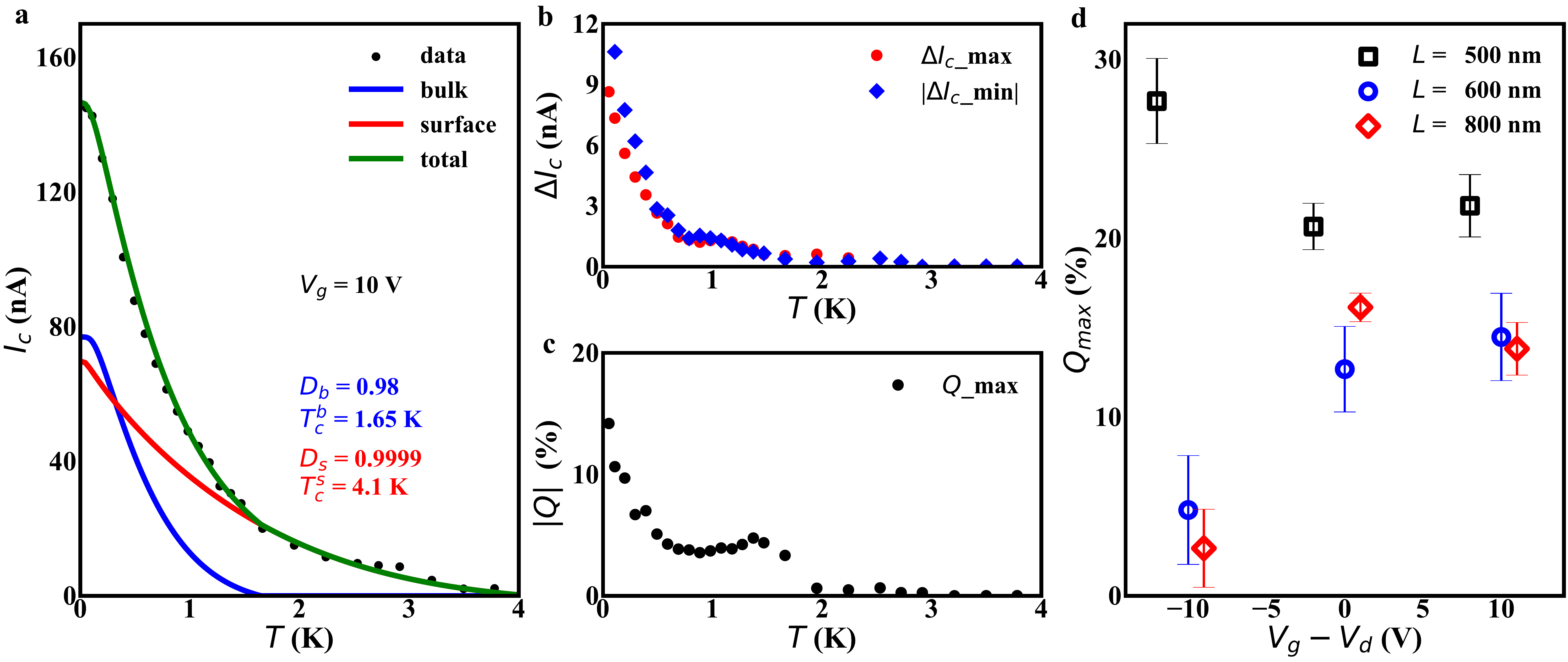}
		\caption{\label{fig:Fig5}
		\textbf{Temperature and length dependence of the JDE (device B: L=600 nm).}  
		\textbf{(a)} Temperature dependence of $I_c$ at $V_g = 10~\mathrm{V}$. The measured $I_c$ (black dots) decreases monotonically with increasing temperature. Two-channel Eilenberger fits are shown as solid lines for the surface contribution (red), bulk contribution (blue), and the total supercurrent (green).  
		\textbf{(b)} Temperature dependence of $\Delta I_c$ corresponding to the maximal (red) and minimal (blue) values under positive and negative magnetic fields, respectively.  
		\textbf{(c)} Temperature dependence of the $Q$-factor, revealing a pronounced enhancement near $1.3~\mathrm{K}$.  
		\textbf{(d)} Gate dependence of the maximal $Q$-factor for junction lengths of 500, 600, and 800~nm.
		}
		
	\end{figure}
	
	\noindent In the previous section, we demonstrated the gate-tunable JDE, with a pronounced enhancement arising from topological surface states. To further distinguish the JDE in the topological regime, we investigate its temperature dependence. Figure~\ref{fig:Fig5}a shows the critical current as a function of the temperature $I_c(T)$ at $V_G=10$ V. The $I_c$ decreases rapidly with temperature without saturation at low $T$. Given the long mean free path ($l_e \sim 600-1000$ nm) \cite{Li2018a}, this behavior is consistent with a ballistic junction regime. Around $T \sim 1.5$ K, a kink emerges in $I_c(T)$, suggesting a two-channel contribution. To capture this behavior, we fit the data using two-channel Eilenberger equations, assigning one channel to bulk states and the other to surface states. 
	
	The fitting parameters include the transparency D and the number of modes N (see SI for details). The superconducting coherence length is estimated as $\xi_s = \frac{\hbar v_F}{\pi \Delta^*}$, where $v_F$ is the Fermi velocity, taken from literature \cite{Li2018a}, and \(\Delta^*\) is the induced superconducting gap beneath the Nb electrodes, determined by the induced superconductivity $T_c^*$. We obtain $\xi_s^{\mathrm{bulk}} \approx 300~\mathrm{nm}$ and $\xi_s^{\mathrm{surf}} \approx 330~\mathrm{nm}$ (see supplementary material for the fitting parameters).
	
	The fit reveals that the bulk channel (blue solid line in Fig.~\ref{fig:Fig5}a) dominates for \(T \lesssim 2~\mathrm{K}\), but its contribution rapidly decreases with increasing temperature due to its ballistic nature. In contrast, the surface channel persists to higher temperatures, consistent with topological protection, and produces the extended ``tail'' in \(I_c(T)\), similar to previous observations in topological insulators \cite{Rosenbach2021}.
	
	Notably, near the crossover temperature, we observe a pronounced enhancement in both \(\Delta I_c\) and the \(Q\)-factor (Fig.~\ref{fig:Fig5}b,c), with a clear peak around 1.3~K. This suggests that as surface states begin to dominate, they contribute more strongly to the JDE than bulk states, leading to increased \(\Delta I_c\) and \(Q\). Such enhancement may serve as a hallmark of topological contributions in a multichannel system.
	
	
	At the end, we present the gate-dependent JDE for junctions with different lengths. In Fig.~\ref{fig:Fig5}d, we plot the maximum Q-factor of three junctions (with L = 500, 600, and 800~nm) as a function of relative gate voltage with respect to the Dirac point ($V_D$). We find that the shortest junction (L = 500~nm) exhibits an overall larger Q-factor and does not show a maximum near $V_g = 10$~V, whereas the 600~nm and 800~nm junctions display very similar gate dependence, both peaking close to the Dirac point. This behavior highlights the importance of higher-harmonic contributions in the JDE: in shorter junctions, higher harmonics can participate more effectively, thereby enhancing the observed diode effect. The fact that the maximum is not present at the same gate voltage may be due to a higher electron-doping level in the junction when the length is short.
	
	\FloatBarrier
	\section{Conclusion}
    In conclusion, we demonstrate a gate-tunable Josephson diode effect in Cd$_3$As$_2$-based Josephson junctions. By establishing a comprehensive model for the amplitude-dependent anisotropic JDE, we provide a clear framework for the rational design of devices with tailored diode anisotropy. Our results reveal a pronounced enhancement of the JDE in both its temperature and length dependence, highlighting the non-trivial interplay between topological surface states and bulk transport. These findings establish the Josephson diode effect as a powerful, symmetry-sensitive probe of topological states and pave the way toward novel superconducting devices and detection schemes based on topological quantum materials.

    \section{Method}
    \subsection*{Material growth}
    \noindent The Cd$_3$As$_2$ nanowires were prepared by chemical vapour deposition method in tube furnace. Cd$_3$As$_2$ powders were placed at the centre of the furnace and silicon wafers coated with a thin gold film about 5 nm in thickness were used as substrates to collect the products downstream. First the tube furnace was flushed several times with Argon gas to fully remove oxygen. Then the temperature was gradually increased to 650 \textdegree C and kept for 10 min with an Argon flow of 20 s.c.c.m. as carrier gas. After the growth process, the furnace was cooled naturally. 

    \subsection*{Device fabrication}
    \noindent Cd$_3$As$_2$ nanowires were first transferred onto a substrate and patterned using standard electron-beam lithography. Subsequently, Nb/Pd (100/5 nm) contacts were deposited to serve as superconducting leads.

	\FloatBarrier
	\bibliography{references}

@article{Cayao2024,
	title = {Enhancing the Josephson diode effect with Majorana bound states},
	author = {Cayao, Jorge and Nagaosa, Naoto and Tanaka, Yukio},
	journal = {Phys. Rev. B},
	volume = {109},
	pages = {L081405},
	year = {2024},
	month = {Feb},
	doi = {10.1103/PhysRevB.109.L081405}
}

@article{Picoli2023,
	title = {Superconducting diode effect in quasi-one-dimensional systems},
	author = {de Picoli, Tatiana and Blood, Zane and Lyanda-Geller, Yuli and V\"ayrynen, Jukka I.},
	journal = {Phys. Rev. B},
	volume = {107},
	pages = {224518},
	year = {2023},
	month = {Jun},
	doi = {10.1103/PhysRevB.107.224518}
}

@article{Mondal2025,
	title = {Josephson diode effect with Andreev and Majorana bound states},
	author = {Mondal, Sayan and Fu, Po-Hao and Cayao, Jorge},
	journal = {arXiv preprint arXiv:2503.08318},
	year = {2025},
	note = {Submitted for publication},
	url = {https://arxiv.org/abs/2503.08318}
}

@article{Wang2024,
	title = {Efficient Josephson diode effect on a two-dimensional topological insulator with asymmetric magnetization},
	author = {Wang, J. and Jiang, Y. and Wang, Juan Juan and Liu, Jun-Feng},
	journal = {Phys. Rev. B},
	volume = {109},
	issue = {7},
	pages = {075412},
	numpages = {7},
	year = {2024},
	month = {Feb},
	publisher = {American Physical Society},
	doi = {10.1103/PhysRevB.109.075412},
	url = {https://link.aps.org/doi/10.1103/PhysRevB.109.075412}
}

@article{Ando2020,
	author = {Ando, Fumihiro and Miyasaka, Yuta and Li, Tian and Ishizuka, Jun and Arakawa, Tomonori and Shiota, Yoichi and Moriyama, Takahiro and Yanase, Youichi and Ono, Teruo},
	title = {Observation of superconducting diode effect},
	journal = {Nature},
	year = {2020},
	volume = {584},
	pages = {373--376},
	doi = {10.1038/s41586-020-2590-4},
	url = {https://www.nature.com/articles/s41586-020-2590-4}
}

@article{Nadeem2023,
	author = {Nadeem, Muhammad and Fuhrer, Michael S. and Wang, Xiaolin},
	title = {The superconducting diode effect},
	journal = {Nature Reviews Physics},
	year = {2023},
	volume = {5},
	pages = {558--576},
	doi = {10.1038/s42254-023-00519-5},
	url = {https://www.nature.com/articles/s42254-023-00519-5}
}

@article{Daido2022,
	author = {Daido, Akito and Ikeda, Yuhei and Yanase, Youichi},
	title = {Intrinsic superconducting diode effect},
	journal = {Physical Review Letters},
	year = {2022},
	volume = {128},
	number = {3},
	pages = {037001},
	doi = {10.1103/PhysRevLett.128.037001},
	url = {https://link.aps.org/doi/10.1103/PhysRevLett.128.037001}
}

@article{Davydova2022_Univ,
	author = {Margarita Davydova and Saranesh Prembabu and Liang Fu},
	doi = {10.1126/SCIADV.ABO0309},
	issn = {23752548},
	issue = {23},
	journal = {Science Advances},
	month = {6},
	pages = {309},
	pmid = {35675396},
	publisher = {American Association for the Advancement of Science},
	title = {Universal Josephson diode effect},
	volume = {8},
	url = {/doi/pdf/10.1126/sciadv.abo0309?download=true},
	year = {2022}
}

@article{Zazunov2024,
	author = {A. Zazunov and J. Rech and T. Jonckheere and B. Grémaud and T. Martin and R. Egger},
	doi = {10.1103/PhysRevB.109.024504},
	issn = {24699969},
	issue = {2},
	journal = {Physical Review B},
	month = {1},
	publisher = {American Physical Society},
	title = {Nonreciprocal charge transport and subharmonic structure in voltage-biased Josephson diodes},
	volume = {109},
	year = {2024}
}

@article{Dolcini2015_phi,
	author  = {Dolcini, Fabrizio and Houzet, Manuel and Meyer, Julia S.},
	title   = {Topological Josephson $\phi_0$ junctions},
	journal = {Phys. Rev. B},
	volume  = {92},
	number  = {3},
	pages   = {035428},
	year    = {2015},
	doi     = {10.1103/PhysRevB.92.035428}
}

@article{Yokoyama2014_phi,
	author = {Tomohiro Yokoyama and Mikio Eto and Yuli V. Nazarov},
	doi = {10.1103/PhysRevB.89.195407},
	issn = {1550235X},
	issue = {19},
	journal = {Physical Review B - Condensed Matter and Materials Physics},
	month = {5},
	publisher = {American Physical Society},
	title = {Anomalous Josephson effect induced by spin-orbit interaction and Zeeman effect in semiconductor nanowires},
	volume = {89},
	year = {2014}
}

@article{Fu2008_MZM,
	title = {Superconducting Proximity Effect and Majorana Fermions at the Surface of a Topological Insulator},
	author = {Fu, Liang and Kane, C. L.},
	journal = {Phys. Rev. Lett.},
	volume = {100},
	issue = {9},
	pages = {096407},
	numpages = {4},
	year = {2008},
	month = {Mar},
	publisher = {American Physical Society},
	doi = {10.1103/PhysRevLett.100.096407},
	url = {https://link.aps.org/doi/10.1103/PhysRevLett.100.096407}
}

@article{Linder2010_MZM,
	title = {Unconventional Superconductivity on a Topological Insulator},
	author = {Linder, Jacob and Tanaka, Yukio and Yokoyama, Takehito and Sudb\o{}, Asle and Nagaosa, Naoto},
	journal = {Phys. Rev. Lett.},
	volume = {104},
	issue = {6},
	pages = {067001},
	numpages = {4},
	year = {2010},
	month = {Feb},
	publisher = {American Physical Society},
	doi = {10.1103/PhysRevLett.104.067001},
	url = {https://link.aps.org/doi/10.1103/PhysRevLett.104.067001}
}

@article{Legg2023_Parity,
	author = {Henry F. Legg and Katharina Laubscher and Daniel Loss and Jelena Klinovaja},
	doi = {10.1103/PhysRevB.108.214520},
	issn = {24699969},
	issue = {21},
	journal = {Physical Review B},
	pages = {1-7},
	publisher = {American Physical Society},
	title = {Parity-protected superconducting diode effect in topological Josephson junctions},
	volume = {108},
	year = {2023}
}

@article{Pal2022,
	author = {Pal, Banabir and Chakraborty, Anirban and Sivakumar, Pranava K. and Davydova, Margarita and Gopi, Ajesh K. and Pandeya, Avanindra K. and Krieger, Jonas A. and Zhang, Yang and Date, Mihir and Ju, Sailong and Yuan, Noah and Schröter, Niels B. M. and Fu, Liang and Parkin, Stuart S. P.},
	title = {Josephson diode effect from Cooper pair momentum in a topological semimetal},
	journal = {Nature Physics},
	year = {2022},
	volume = {18},
	pages = {277--282},
	doi = {10.1038/s41567-022-01699-5},
	url = {https://www.nature.com/articles/s41567-022-01699-5}
}

@article{ZhangPRX2022,
	author = {Zhang, Yi and Gu, Yuhao and Li, Pengfei and Hu, Jiangping and Jiang, Kun},
	title = {General theory of Josephson diodes},
	journal = {Physical Review X},
	year = {2022},
	volume = {12},
	number = {4},
	pages = {041013},
	doi = {10.1103/PhysRevX.12.041013},
	url = {https://link.aps.org/doi/10.1103/PhysRevX.12.041013}
}

@article{Lu2023,
	author = {Lu, Bo and Ikegaya, Satoshi and Burset, Pablo and Tanaka, Yukio and Nagaosa, Naoto},
	title = {Tunable Josephson diode effect on the surface of topological insulators},
	journal = {Physical Review Letters},
	year = {2023},
	volume = {131},
	number = {9},
	pages = {096001},
	doi = {10.1103/PhysRevLett.131.096001},
	url = {https://link.aps.org/doi/10.1103/PhysRevLett.131.096001}
}

@article{Tanaka2022,
	author = {Tanaka, Yukio and Lu, Bo and Nagaosa, Naoto},
	doi = {10.1103/PhysRevB.106.214524},
	issn = {24699969},
	journal = {Physical Review B},
	number = {21},
	pages = {1--13},
	title = {{Theory of giant diode effect in d -wave superconductor junctions on the surface of a topological insulator}},
	volume = {106},
	year = {2022}
}

@article{Wu2022_Nb3Br8,
	author = {Heng Wu and Yaojia Wang and Yuanfeng Xu and Pranava K. Sivakumar and Chris Pasco and Ulderico Filippozzi and Stuart S.P. Parkin and Yu Jia Zeng and Tyrel McQueen and Mazhar N. Ali},
	doi = {10.1038/s41586-022-04504-8},
	issn = {1476-4687},
	issue = {7907},
	journal = {Nature},
	month = {4},
	pages = {653-656},
	pmid = {35478238},
	publisher = {Nature Publishing Group},
	title = {The field-free Josephson diode in a van der Waals heterostructure},
	volume = {604},
	url = {https://www.nature.com/articles/s41586-022-04504-8},
	year = {2022}
}

@article{Bauriedl2022_NbSe2,
	author = {Lorenz Bauriedl and Christian Bäuml and Lorenz Fuchs and Christian Baumgartner and Nicolas Paulik and Jonas M. Bauer and Kai Qiang Lin and John M. Lupton and Takashi Taniguchi and Kenji Watanabe and Christoph Strunk and Nicola Paradiso},
	doi = {10.1038/s41467-022-31954-5},
	issn = {2041-1723},
	issue = {1},
	journal = {Nature Communications},
	month = {7},
	pages = {4266-},
	pmid = {35871226},
	publisher = {Nature Publishing Group},
	title = {Supercurrent diode effect and magnetochiral anisotropy in few-layer NbSe2},
	volume = {13},
	url = {https://www.nature.com/articles/s41467-022-31954-5},
	year = {2022}
}

@article{deVries2021_Gra,
	author = {Folkert K. de Vries and Elías Portolés and Giulia Zheng and Takashi Taniguchi and Kenji Watanabe and Thomas Ihn and Klaus Ensslin and Peter Rickhaus},
	doi = {10.1038/s41565-021-00896-2},
	issn = {1748-3395},
	issue = {7},
	journal = {Nature Nanotechnology},
	month = {5},
	pages = {760-763},
	pmid = {33941917},
	publisher = {Nature Publishing Group},
	title = {Gate-defined Josephson junctions in magic-angle twisted bilayer graphene},
	volume = {16},
	url = {https://www.nature.com/articles/s41565-021-00896-2},
	year = {2021}
}

@article{Dez-Mrida2023_Gra,
	author = {J. Díez-Mérida and A. Díez-Carlón and S. Y. Yang and Y. M. Xie and X. J. Gao and J. Senior and K. Watanabe and T. Taniguchi and X. Lu and A. P. Higginbotham and K. T. Law and Dmitri K. Efetov},
	doi = {10.1038/s41467-023-38005-7},
	issn = {2041-1723},
	issue = {1},
	journal = {Nature Communications 2023 14:1},
	month = {4},
	pages = {2396-},
	pmid = {37100775},
	publisher = {Nature Publishing Group},
	title = {Symmetry-broken Josephson junctions and superconducting diodes in magic-angle twisted bilayer graphene},
	volume = {14},
	url = {https://www.nature.com/articles/s41467-023-38005-7},
	year = {2023}
}

@article{Lin2022_Gra,
	author = {Jiang Xiazi Lin and Phum Siriviboon and Harley D. Scammell and Song Liu and Daniel Rhodes and K. Watanabe and T. Taniguchi and James Hone and Mathias S. Scheurer and J. I.A. Li},
	doi = {10.1038/s41567-022-01700-1},
	issn = {1745-2481},
	issue = {10},
	journal = {Nature Physics},
	month = {8},
	pages = {1221-1227},
	publisher = {Nature Publishing Group},
	title = {Zero-field superconducting diode effect in small-twist-angle trilayer graphene},
	volume = {18},
	url = {https://www.nature.com/articles/s41567-022-01700-1},
	year = {2022}
}

@article{Pal2022_NiTe2,
	author = {Banabir Pal and Anirban Chakraborty and Pranava K. Sivakumar and Margarita Davydova and Ajesh K. Gopi and Avanindra K. Pandeya and Jonas A. Krieger and Yang Zhang and Mihir Date and Sailong Ju and Noah Yuan and Niels B.M. Schröter and Liang Fu and Stuart S.P. Parkin},
	doi = {10.1038/s41567-022-01699-5},
	isbn = {4156702201699},
	issn = {17452481},
	issue = {10},
	journal = {Nature Physics},
	month = {10},
	pages = {1228-1233},
	publisher = {Nature Research},
	title = {Josephson diode effect from Cooper pair momentum in a topological semimetal},
	volume = {18},
	year = {2022}
}

@article{Chen2024_MoTe2,
	author = {Pingbo Chen and Gongqi Wang and Bicong Ye and Jinhua Wang and Liang Zhou and Zhenzhong Tang and Le Wang and Jiannong Wang and Wenqing Zhang and Jiawei Mei and Weiqiang Chen and Hongtao He},
	doi = {10.1002/ADFM.202311229;REQUESTEDJOURNAL:JOURNAL:16163028;WGROUP:STRING:PUBLICATION},
	issn = {16163028},
	issue = {10},
	journal = {Advanced Functional Materials},
	month = {3},
	pages = {2311229},
	publisher = {John Wiley and Sons Inc},
	title = {Edelstein Effect Induced Superconducting Diode Effect in Inversion Symmetry Breaking MoTe2 Josephson Junctions},
	volume = {34},
	year = {2024}
}

@article{Baumgartner2022,
	author = {Baumgartner, Christian and Fuchs, Lorenz and Costa, Andreas and Reinhardt, Simon and Gronin, Sergei and Gardner, Geoffrey C. and Lindemann, Tyler and Manfra, Michael J. and {Faria Junior}, Paulo E. and Kochan, Denis and Fabian, Jaroslav and Paradiso, Nicola and Strunk, Christoph},
	doi = {10.1038/s41565-021-01009-9},
	issn = {17483395},
	journal = {Nature Nanotechnology},
	number = {1},
	pages = {39--44}, 
	publisher = {Springer US},
	title = {{Supercurrent rectification and magnetochiral effects in symmetric Josephson junctions}},
	volume = {17},
	year = {2022}
}

@article{Turini2022_InSb,
	author = {Bianca Turini and Sedighe Salimian and Matteo Carrega and Andrea Iorio and Elia Strambini and Francesco Giazotto and Valentina Zannier and Lucia Sorba and Stefan Heun},
	doi = {10.1021/ACS.NANOLETT.2C02899},
	issn = {15306992},
	issue = {21},
	journal = {Nano Letters},
	month = {11},
	pages = {8502-8508},
	pmid = {36285780},
	publisher = {American Chemical Society},
	title = {Josephson Diode Effect in High-Mobility InSb Nanoflags},
	volume = {22},
	year = {2022}
}

@article{Li2018a,
	author = {Li, Cai Zhen and Li, Chuan and Wang, Li Xian and Wang, Shuo and Liao, Zhi Min and Brinkman, Alexander and Yu, Da Peng},
	doi = {10.1103/PhysRevB.97.115446},
	journal = {Physical Review B},
	number = {11},
	pages = {115446},
	title = {{Bulk and surface states carried supercurrent in ballistic Nb-Dirac semimetal Cd3As2 nanowire-Nb junctions}},
	url = {https://doi.org/10.1103/PhysRevB.97.115446},
	volume = {97},
	year = {2018}
}

@article{Li2021b,
	author = {Li, Cai Zhen and Wang, An Qi and Li, Chuan and Zheng, Wen Zhuang and Brinkman, Alexander and Yu, Da Peng and Liao, Zhi Min},
	doi = {10.1103/PHYSREVLETT.126.027001/FIGURES/4/MEDIUM},
	issn = {10797114},
	journal = {Physical Review Letters},
	month = {jan},
	number = {2},
	pages = {027001},
	publisher = {American Physical Society},
	title = {{Topological Transition of Superconductivity in Dirac Semimetal Nanowire Josephson Junctions}},
	url = {https://journals.aps.org/prl/abstract/10.1103/PhysRevLett.126.027001},
	volume = {126},
	year = {2021}
}

@article{Li2016,
	author = {Li, Caizhen and Jia, Zhenzhao and Li, Xinqi and Shi, Junren and Liao, Zhimin and Yu, Dapeng and Wu, Xiaosong},
	title = {Aharonov-Bohm oscillations in Dirac semimetal Cd3As2 nanowires},
	journal = {Nature Communications},
	volume = {7},
	pages = {10769},
	year = {2016},
	month = {Feb},
	publisher = {Nature Publishing Group},
	doi = {10.1038/ncomms10769},
	url = {https://doi.org/10.1038/ncomms10769}
}

@article{Ali2014,
	author = {Ali, Mazhar N. and Gibson, Quinn and Jeon, Sangjun and Zhou, Brian B. and Yazdani, Ali and Cava, R. J.},
	title = {The Crystal and Electronic Structures of Cd3As2, the Three-Dimensional Electronic Analogue of Graphene},
	journal = {Inorganic Chemistry},
	volume = {53},
	number = {8},
	pages = {4062-4067},
	year = {2014},
	doi = {10.1021/ic403163d},
	URL = {  https://doi.org/10.1021/ic403163d},
	
}

@article{Crassee2018,
	author    = {Crassee, Ilari and Armitage, N. P.},
	title     = {3D Dirac semimetals: a review of material properties},
	journal   = {Physical Review Materials},
	year      = {2018},
	volume    = {2},
	number    = {12},
	pages     = {120302},
	doi       = {10.1103/PhysRevMaterials.2.120302},
	url       = {https://journals.aps.org/prmaterials/abstract/10.1103/PhysRevMaterials.2.120302},
	note      = {Review article}
}

@article{Li2020e,
	author = {Li, Cai Zhen and Wang, An Qi and Li, Chuan and Zheng, Wen Zhuang and Brinkman, Alexander and Yu, Da Peng and Liao, Zhi Min},
	doi = {10.1038/s41467-020-15010-8},
	issn = {2041-1723},
	journal = {Nature Communications},
	month = {mar},
	number = {1},
	pages = {1--7},
	publisher = {Nature Publishing Group},
	title = {{Fermi-arc supercurrent oscillations in Dirac semimetal Josephson junctions}},
	url = {https://www.nature.com/articles/s41467-020-15010-8},
	volume = {11},
	year = {2020}
}

@article{Cao2015,
	author = {Junzhi Cao and Sihang Liang and Cheng Zhang and Yanwen Liu and Junwei Huang and Zhao Jin and Zhi Gang Chen and Zhijun Wang and Qisi Wang and Jun Zhao and Shiyan Li and Xi Dai and Jin Zou and Zhengcai Xia and Liang Li and Faxian Xiu},
	doi = {10.1038/ncomms8779},
	issn = {2041-1723},
	issue = {1},
	journal = {Nature Communications 2015 6:1},
	month = {7},
	pages = {1-6},
	publisher = {Nature Publishing Group},
	title = {Landau level splitting in Cd3As2 under high magnetic fields},
	volume = {6},
	url = {https://www.nature.com/articles/ncomms8779},
	year = {2015}
}

@article{Liu2014,
	author = {Z. K. Liu and J. Jiang and B. Zhou and Z. J. Wang and Y. Zhang and H. M. Weng and D. Prabhakaran and S. K. Mo and H. Peng and P. Dudin and T. Kim and M. Hoesch and Z. Fang and X. Dai and Z. X. Shen and D. L. Feng and Z. Hussain and Y. L. Chen},
	doi = {10.1038/nmat3990},
	issn = {1476-4660},
	issue = {7},
	journal = {Nature Materials},
	month = {5},
	pages = {677-681},
	title = {A stable three-dimensional topological Dirac semimetal Cd3As2},
	volume = {13},
	url = {https://www.nature.com/articles/nmat3990},
	year = {2014}
}

@article{Jeon2014,
	author = {Sangjun Jeon and Brian B. Zhou and Andras Gyenis and Benjamin E. Feldman and Itamar Kimchi and Andrew C. Potter and Quinn D. Gibson and Robert J. Cava and Ashvin Vishwanath and Ali Yazdani},
	doi = {10.1038/nmat4023},
	issn = {1476-4660},
	issue = {9},
	journal = {Nature Materials 2014 13:9},
	month = {6},
	pages = {851-856},
	title = {Landau quantization and quasiparticle interference in the three-dimensional Dirac semimetal Cd3As2},
	volume = {13},
	url = {https://www.nature.com/articles/nmat4023},
	year = {2014}
}

@article{Rosenbach2021,
	author = {Rosenbach, Daniel and Schmitt, Tobias W. and Sch{\"{u}}ffelgen, Peter and Stehno, Martin P. and Li, Chuan and Schleenvoigt, Michael and Jalil, Abdur R. and Mussler, Gregor and Neumann, Elmar and Trellenkamp, Stefan and Golubov, Alexander A. and Brinkman, Alexander and Gr{\"{u}}tzmacher, Detlev and Sch{\"{a}}pers, Thomas},
	doi = {10.1126/sciadv.abf1854},
	journal = {Science Advances},
	month = {jun},
	number = {26},
	pages = {1--10},
	pmid = {34162537},
	publisher = {American Association for the Advancement of Science},
	title = {{Reappearance of first Shapiro step in narrow topological Josephson junctions}},
	volume = {7},
	year = {2021}
}

	
\end{document}




\renewcommand{\thefigure}{S\arabic{figure}}
\begin{figure}[hpt]
    \includegraphics[width=1\columnwidth]{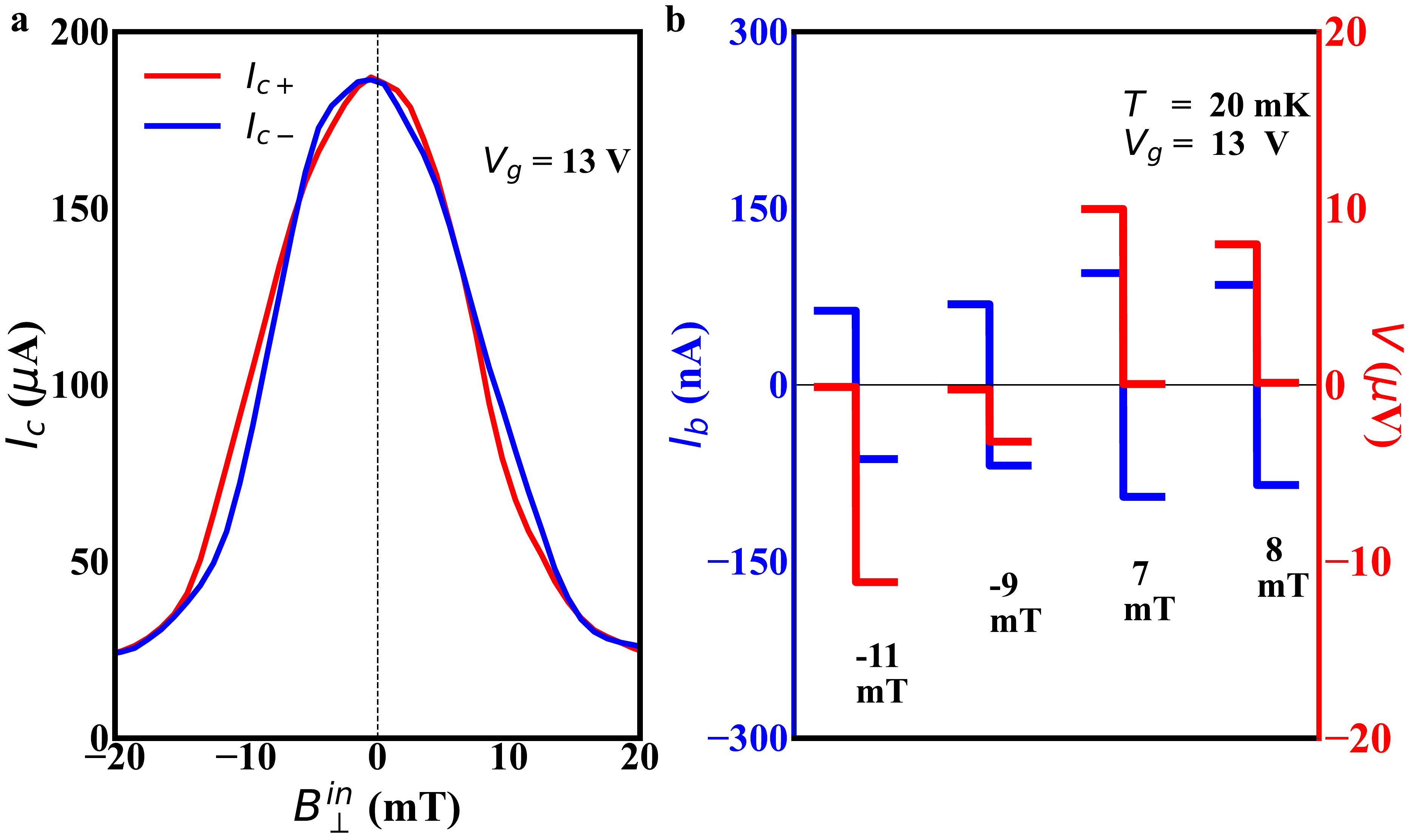}
    \captionsetup{font={footnotesize,stretch=1.25},labelfont={footnotesize,bf}}
    \caption{\label{fig:SI_Fig1} 
    Josephson diode effect at 20 mK(device B: L=600 nm). \textbf{(a)} The switching supercurrent $I^+_c$ and $\lvert I^-_c\rvert$ as a function of $B_{\perp}^{in}$ at $V_g$=13 V. It shows a largest asymmetry in switching supercurrent at $B_{\perp}^{in}$ $\sim \pm$11 mT.\textbf{(b)}  Observed rectification effect in voltage across junction by using absolute value of bias current between $I^+_c$ and $I^-_c$ at $T$= 20 mK and under several selected values of $B_{\perp}^{in}$.}
\end{figure}

\begin{figure}[hpt]
    \includegraphics[width=1\columnwidth]{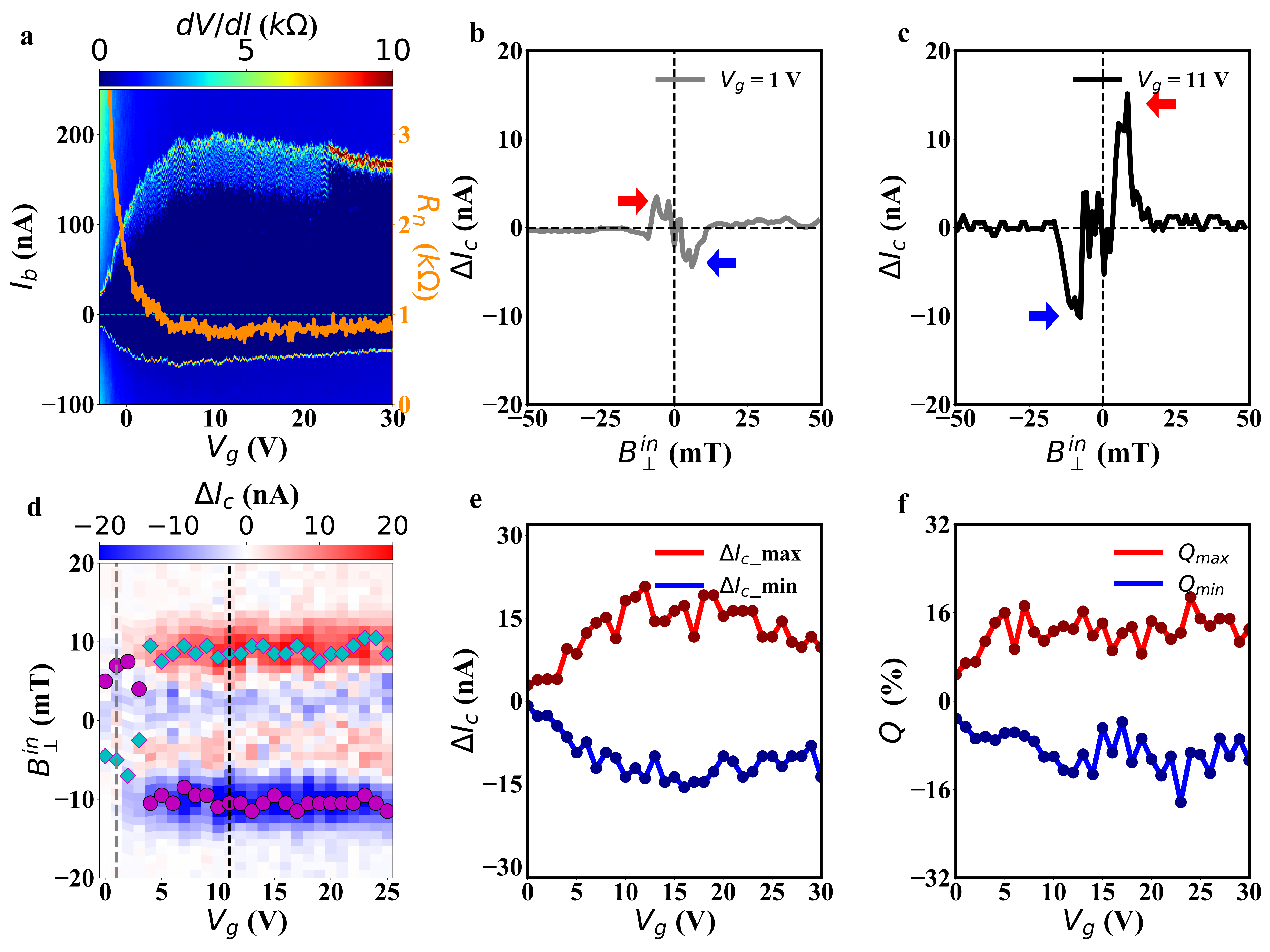}
    \captionsetup{font={footnotesize,stretch=1.25},labelfont={footnotesize,bf}}
    \caption{\label{fig:SI_Fig2} 
    Gate voltage dependence of the Josephson diode effect at 20 mK(device B: L=600 nm). \textbf{(a)} Differential Resistance $dVdI$ as a function of applied gate voltage $V_g$ and bias current $I_b$. It indicates that the critical supercurrent can be largely suppressed by the negative gate voltage. \textbf{(b,c)} Dependence of $\Delta$$I_c$ on in-plane perpendicular magnetic field $B_{\perp}^{in}$ under $V_g$=1 V and $V_g$=11 V, respectively. Arrow in red shows the position of the curve's maximum point, while the arrow in blue indicates the position of the curve's minimum point. \textbf{(d)} $\Delta$$I_c$ as a function of applied gate voltage and in-plane perpendicular magnetic field. Diamond-shaped points in cyan mark the position of maximum value, while circular dot mark the position of minimum value. \textbf{(e,f)} $\Delta I_{c_-}$max, $\Delta I_{c_-}$min, $Q_{max}$, and $Q_{min}$ as a function of $V_g$.}
\end{figure}

\begin{figure}[hpt]
    \includegraphics[width=1\columnwidth]{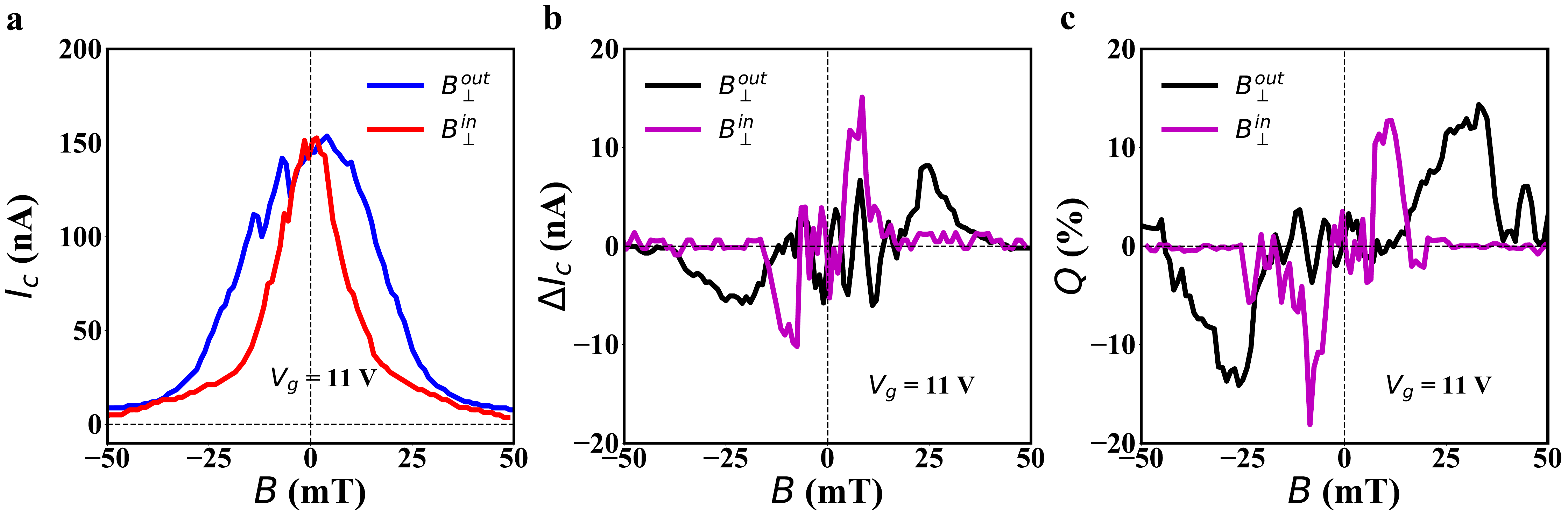}
    \captionsetup{font={footnotesize,stretch=1.25},labelfont={footnotesize,bf}}
    \caption{\label{fig:SI_Fig3} 
    Angle-dependence of JDE in the plane that is perpedicular to the wire(device B: L=600 nm). \textbf{(a)} Dependence of critical supercurrent $I_c$ on $B_{\perp}^{out}$ and $B_{\perp}^{in}$, respectively. \textbf{(b)} $\Delta$$I_c$ as a function of $B_{\perp}^{out}$ and  $B_{\perp}^{in}$. \textbf{(c)} $Q$ as a function of $B_{\perp}^{out}$ and  $B_{\perp}^{in}$.}
\end{figure}

\renewcommand{\figurename}{SI-Fig}
\begin{figure}[hpt]
    \includegraphics[width=1\columnwidth]{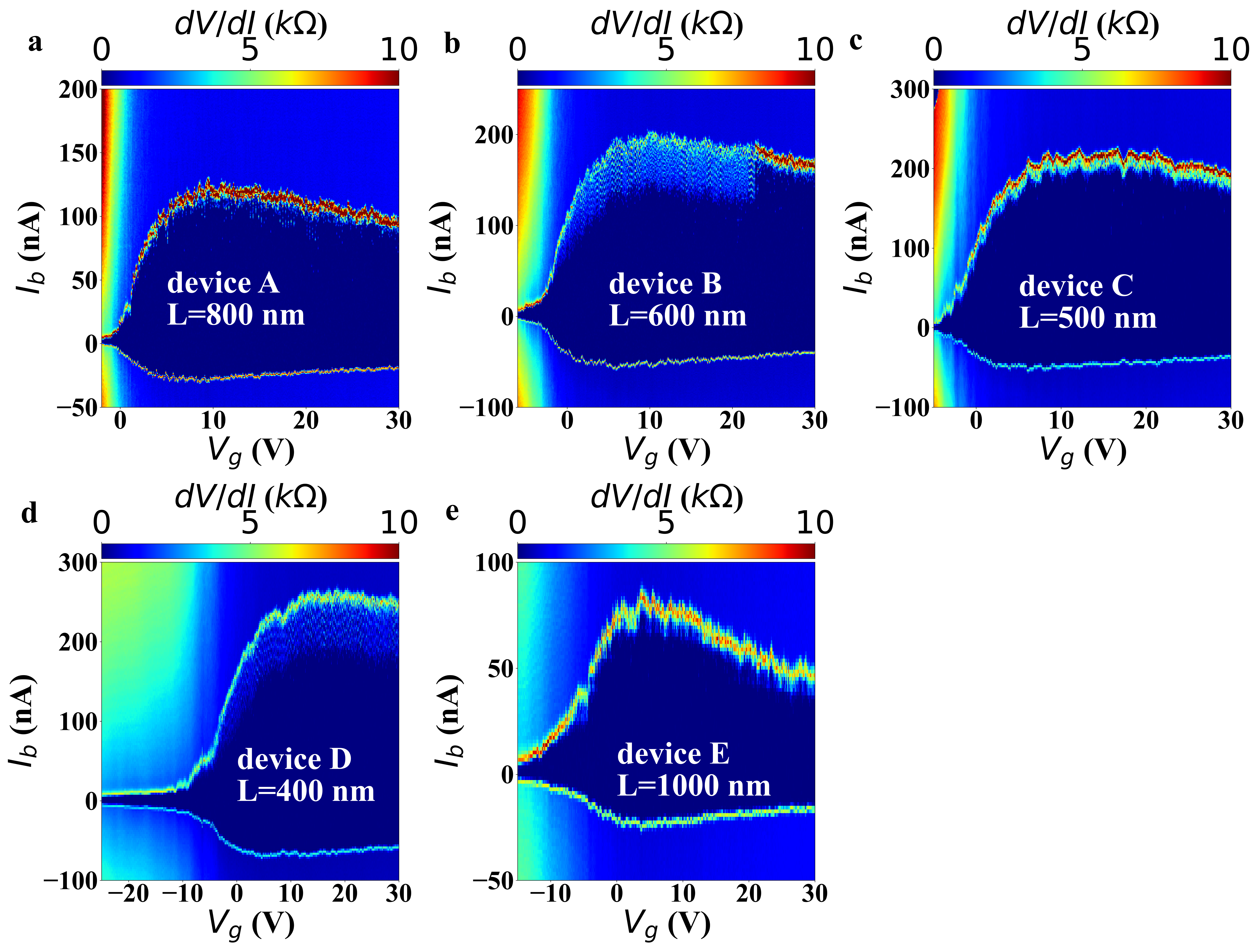}
    \captionsetup{font={footnotesize,stretch=1.25},labelfont={footnotesize,bf}}
    \caption{\label{fig:SI_Fig4} 
    \textbf{(a-e)} Differential Resistance $dV/dI$ as a function of applied gate voltage $V_g$ and bias current $I_b$ for device A-E. Length: L=400 nm, L=500 nm, L=600 nm, L=800 nm and L=1000 nm, respectively. }
\end{figure}

\begin{figure}[hpt]
    \includegraphics[width=1\columnwidth]{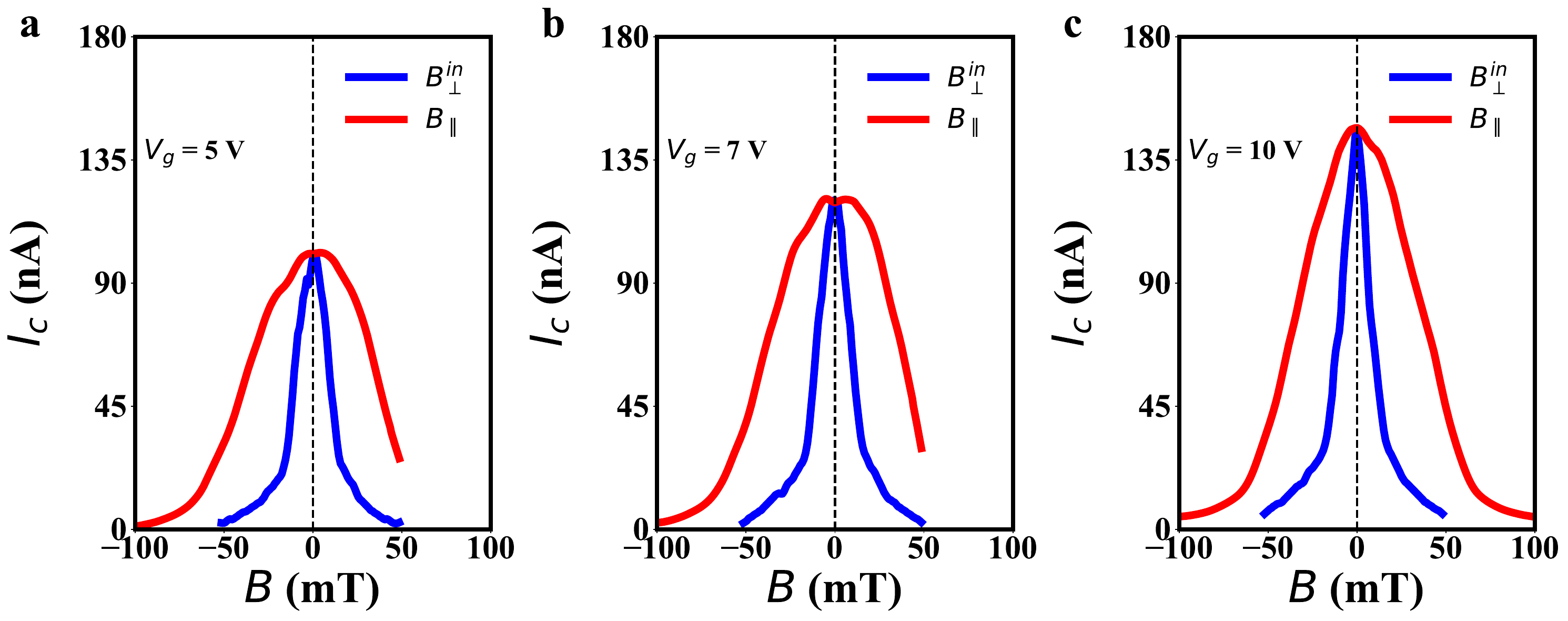}
    \captionsetup{font={footnotesize,stretch=1.25},labelfont={footnotesize,bf}}
    \caption{\label{fig:SI_Fig4} 
    \textbf{(a-c)} device B(L=600 nm) : Both $I_c$ as a function of applied in-plane perpedicular magnetic field and $I_c$ as a function of applied in-plane parallel magnetic field at Vg=5V, 7V and 10V, respectively.}
\end{figure}

\begin{table}[htbp]
    \centering
    \caption{Parameters for two-channel Eilenberger fit}
    \label{tab:bulk_surface_params}
    \begin{tabular}{lccccccc}
        \toprule
        State & $T_c$ (K) & $\xi$ (nm) & $\Delta^*$ ($\mu$eV) & $v_F$ (m/s) & $R_n$ ($\Omega$) & $N$ (channels) \\
        \midrule
        Bulk states    & 1.65 & 300 & 250.4 & $3.58\times 10^{5}$ & 890  & 15 \\
        Surface states & 4.10 & 330 & 622.3 & $9.79\times 10^{5}$ & 3600 &  4 \\
        \bottomrule
    \end{tabular}
\end{table}